\documentclass{article}
\usepackage{graphics}
\usepackage{latexsym}

\newcommand{\be}{\begin{equation}}
\newcommand{\R}{\mathbf{R}}
\newcommand{\bw}{\mathbf{w}}
\newcommand{\ee}{\end{equation}}
\newcommand{\z}{\tilde}

\title{Fokker-Planck Equation on Fractal Curves}
\author{Seema Satin$^{1,2}$ \and  Abhay Parvate$^{1,2}$ \and A.D.Gangal$^1$\\ 
\and $1$ Department of Physics, University of Pune, Pune 411~007, India\\
\and $2$ Center for Modelling and Simulation, University of Pune, Pune 411~007,
India \\
\and \texttt{satin@cms.unipune.ernet.in},
\and \texttt{abhay@physics.unipune.ernet.in},
\and \texttt{adg@physics.unipune.ernet.in}
}
\date{ }
\begin{document}
\maketitle
\begin{abstract}
 A Fokker-Planck equation on fractal curves is obtained,
 starting from Chapmann-Kolmogorov equation on fractal curves. This is done
using the recently developed calculus on fractals, which allows one to
write differential equations on fractal curves. As an
important special case, the diffusion and drift coefficients are obtained, for
 a suitable transition probability to get the diffusion equation on fractal
 curves.
This equation is of first order in time, and, in space variable it involves
derivatives of order $\alpha$, $\alpha$ being the dimension of the curve. 
The solution of this equation with localized initial condition shows
 deviation from ordinary diffusion behaviour due to underlying fractal space
 in which diffusion is taking place. An exact solution of this equation 
manifests a subdiffusive behaviour. The dimension of the fractal  path can be
estimated from the distribution function.
\end{abstract}
Fractal curves and paths are encountered frequently in Physics 
\cite{Mandelbrot,Falconer}. 
Several geometries like that of polymer chains, percolating clusters, 
Brownian and Fractional Brownian trajectories and many
 more have been identified as fractals, more precisely as fractal curves 
\cite{Mandelbrot}. Transport on these structures
 reveal remarkable properties \cite{Strichartz,Barlow,Barma,Dhar,Klages,Adv1,
Adv2}. In particular, anomalous diffusion on fractals is a topic of immense 
current interest \cite{Klafter1,Klafter2,Klafter3,Zaslavsky,Zaslavsky1,
Zaslavsky2,Ben,West,Kiran,ben}. There 
have been several investigations \cite{ben1,dyer,Weitkamp,Dozier}, both 
analytical as well as numerical which shed light on the various facets of the 
problem. These investigations use the Fractional operators .
The Fractional Derivatives are often used to explore the 
charcateristic features of anomalous diffusion by setting up fractional kinetic
 equations. Fractional derivatives are non local operators and hence
 not always suitable to handle local scaling behaviour. Analysis on fractals 
is another remarkable development which has been extensively used for the 
treatment of diffusion, heat conduction, waves etc on fractals. But, in this 
approach the operators are constructed using the self similarity of fractal 
sets, and these are restricted to such sets which are post critically finite.
However, as is well known, an ordinary diffusion equation is 
inadequate to describe anomalous transport, and an exact equation based on an
 appropriately developed calculus is still desired.

Ordinary calculus does not equip us to handle problems such as anomalous
 diffusion, dynamics on fractals, fields of fractally distributed sources etc
by setting up and solving ordinary differential equations. Several authors 
have recognized the need to use fractional derivatives and integrals to
explore the characteristic features of fractal walks, anomalous diffusion,
 transport etc. by setting up fractional kinetic equations, master equations
and so on.

Further measure-theoretical approaches are used which include defining 
derivative as inverse of the integral with respect to a measure and defining 
other operators using derivatives. Even though measure theoretical approach
is elegant, Riemann integration like procedures have their own place. They are
more transparent, constructive and advantageous from algorithmic point of veiw.

 Fractal curves lack the smoothness properties required by 
ordinary calculus.  
Recently, \cite{Seema} a new calculus  on fractal curves, such as the von Koch
 curve, was formulated. In this calculus, a~Riemann-like integral along a
fractal curve $F$ is defined. This integral is called the $F^\alpha$-integral,
 where  $\alpha$ is the dimension of the curve.
 A derivative along the fractal curve, called $F^\alpha$-derivative, is also 
defined. These operators are different from the fractional operators used in
 \cite{Klafter1,Klafter2,Klafter3}, in the sense
 that they are local and a newly defined measure like quantity called the mass
 function (which is algorithmic in nature) is used to define these. The 
order of these operators is exactly equal to the dimension of underlying space
, and thus they reduce to ordinary integral and derivative operators when the 
space is $\mathbf{R}$ and not a fractal. 

Several aspects of this calculus retain much of the simplicity of
 ordinary calculus. In fact a conjugacy between this calculus and ordinary
 calculus on the real line is established. This new calculus which is simple,
 direct and algorithmic can be applied to various physical processes.

Here, using the framework of this calculus, we develope a first-principles
approach to the Fokker-Planck equation on fractal curves.
A particular choice of transition probability then leads to a new form of
 diffusion equation. This equation is of the first order in time and  involves
 application of the $F^\alpha$-derivative with respect to spatial variable 
twice. An exact solution of this equation shows a subdiffusive
 behaviour.

We begin by fixing our notation as in \cite{Seema}.
We consider a (fractal) curve $F \subset \mathbf{R^n}$ which is continuously 
paramatrizable i.e there exists a function
$\mathbf{w}:[a_0,b_0] \rightarrow F \subset \mathbf{R^n}$ which is continuous.
We also assume $\bw$ to be invertible. 
 A subdivision $P_{[a,b]}$ of interval $[a,b], a<b,$ is
a finite set of points $\{a=u_0<u_1,\ldots<u_n = b\}$. 
For $a_0 \leq a <b\leq b_0$ and appropriate $\alpha$ to be chosen, let
\be
\gamma^\alpha (F,a,b) = \lim_{\delta \rightarrow 0}\quad
\inf_{\{P_{[a,b]}:|P| \leq \delta\}}
\sum_{i=0}^{n-1}\frac{|\bw(u_{i+1})-\bw(u_i)|^\alpha}
{\Gamma(\alpha+1)}
\ee
where $|\cdot|$ denotes the Euclidean norm on $\R^n$.  
and $|P| = \max \{u_{i+1} - u_i; i = 0,\dots,n-1 \}$
A new dimension, the 
 $\gamma$-dimension of $F$, which will be denoted by $\dim_\gamma(F)$, is given 
by
\[\dim_\gamma(F) = \inf\{\alpha : \gamma^\alpha(F,a,b) = 0\} = \sup
\{\alpha : \gamma^\alpha(F,a,b) = \infty \}\]

Hereafter, $\alpha$ will be assumed to be equal to $\dim_\gamma (F)$
(thus, $\alpha \geq 1$).

The rise (staircase) function $S_F^\alpha: [a_0,b_0]\rightarrow
\mathbf{R}$ of order $\alpha$ for a set $F$,
is defined as  
\be
S_F^{\alpha}(u) = \left\{ \begin{array}{ll}
	\gamma^{\alpha}(F,p_0,u) & u \geq p_0 \\
	- \gamma^\alpha(F,u,p_0) & u < p_0
		\end{array}
	\right. 
	\label{eq:staircase_function} \ee
where $a_0 \leq p_0 \leq b_0$ is arbitrary but fixed,
and $u \in [a_0,b_0]$.
It is a monotonic function.
We denote a point on the fractal curve $F$ by $\theta = \bw(u)$, and
 define
\[J(\theta) = S_F^\alpha(\bw^{-1} (\theta)), \quad \theta \in F\]

Hereafter we consider only those curves for which $S_F^\alpha$ is \textit{finite} and
 \textit{invertible} on $[a,b]$.
 The $F^\alpha$-derivative of a bounded function
 $f : F \rightarrow \R$ at $\theta \in F$ is defined as
\be (D_F^\alpha f)(\theta)= F \mbox{-}\lim_{\theta' \rightarrow \theta} 
\frac{f(\theta')-f(\theta)}
{J(\theta')-J(\theta)} \label{eq:derivative}\ee
where the $F\mbox{-}\lim$ denotes limit along points of $F$,
if the limit exists.

Let $C(a,b)$ denote the segment $\{\bw(u): u \in [a,b]\}$ of $F$. 
A Riemann-like integral on $F$, called $F^\alpha$-integral,
is also defined \cite{Seema}. It is denoted by
$\int_{C(a,b)} f(\theta) d_F^\alpha \theta. $
 The above mentioned $F^\alpha$-derivative and $F^\alpha$- integral are 
related to each other through the Fundamental Theorems of
Calculus as ``inverses'' of each other \cite{Seema}.

The notion of conjugacy of calculus on fractals and ordinary calculus on the
 real line is very useful. Let $\phi$ denote the map (conjugacy) from the class
 of bounded functions on $F$ to the class of bounded functions on the interval 
$[S_F^\alpha(a_0), S_F^\alpha(b_0)]$defined by
$ \phi[f](S_F^\alpha(u))= f(\mathbf{w}(u))$
Then it follows \cite{Seema}
\be \label{eq:conj}  
\int_{S_F^\alpha(a)}^{S_F^\alpha(b)} g(x) dx = \int_{C(a,b)} f(\theta) 
d_F^\alpha \theta \ee
where $g = \phi[f]$.
Moreover $\phi$ relates derivatives $D_F^\alpha$
with the ordinary derivative~$D$, thus $(D_F^\alpha f)\mathbf{w}(u) = (D g)(u)$.

The Taylor series is given by
\be \label{eq:fractaylor}
h(\theta) = \sum_{n=0}^{\infty} \frac{(J(\theta)-J(\theta'))^n}{n!} 
(D_F^\alpha)^n h(\theta')\ee
provided the bounded function $h$ is $F^\alpha$- differentiable any number of 
times on $C(a,b)$. It is also possible to write a Taylor series with remainder.

Let $\int_{C(a,b)}V(\theta,t)d_F^\alpha \theta$ be the probability that
a particle constrained to move on~$F$ is found in the segment~$C(a,b)$, or in
 other words, $V(\theta, t)$ denotes the 'fractal' probability density
that the particle is found at~$\theta$ at time~$t$.
Let the probability density for transition from a point
$\theta'$ at time $t$ to $\theta$ at time $t+\tau$, be 
denoted by $P(\theta,t+\tau|\theta',t)$. A formalism to analyse similar  
situation in ordinary space is developed in \cite{Risken}, we intend to modify
 the same for the case of fractal curves.

The Chapmann-Kolmogorov equation on fractal curve $F$ can be written in the 
form
\be \label{eq:chapkol}
V(\theta,t+\tau) = \int_{C(a,b)}
P(\mathbf{\theta},t+\tau|\theta',t) V(\theta',
t) d_F^\alpha \theta'
\ee
where $\theta, \theta ' \in F$. 
Let 
 $\Delta \equiv \Delta(\theta,\theta') = J(\theta) - J(\theta')$. The 
intergrand in equation (\ref{eq:chapkol}) is:
\begin{eqnarray*}
P(\theta,t+\tau|\theta',t) V(\theta',t)
&=& P(J^{-1}(J(\theta) -\Delta + \Delta),\\
& & t+\tau|J^{-1}(J(\theta)-\Delta),t)\\
& & {} \times V(J^{-1}(J(\theta)-\Delta),t)
\end{eqnarray*}
The Taylor expansion of this integrand in (\ref{eq:chapkol}) then leads to
\begin{eqnarray}
\lefteqn{V(\theta,t+\tau)- V(\theta,t)} \nonumber\\
& = &\int \sum_{n=1}^{\infty}
\frac{(-1)^n}{n!}\Delta^n(D_F^\alpha|_{\theta})^n
\{ P(J^{-1}(J(\theta)+\Delta), \nonumber \\
& & t+\tau|\theta,t)
V(\theta,t) \} d_F^\alpha \theta' \label{eq:difference}
\end{eqnarray}
While there are other ways to write the Fokker-Planck equation we turn to the
use of conjugacy which reduces the problem to the ordinary case and the 
meaning of moments will be transparent. We use the following explicit notation
for the conjugacy map $\phi$ :
\[(\phi_{\theta} H)(J(\theta),\theta') = H(\theta,\theta') \]
\[ (\phi_{\theta'} H)(\theta,J(\theta')) = H(\theta,\theta')\]
while applying to a function $H(\theta,\theta')$ of two arguments $\theta$ and $\theta'$.

Let $y=J(\theta),y'=J(\theta') $ and assume $S_F^\alpha(a_0) \leq y,y' \leq 
S_F^\alpha(b_0)$. Further let us denote
\[\z{\Delta}(\theta,y')= \phi_{\theta'} \Delta(\theta,\theta') \] 
\[\Delta'(y,y')= \phi_{\theta} \z{\Delta} = \phi_{\theta} \circ \phi_{\theta'} 
\Delta(\theta,\theta') \]
Then,
\[\z{\Delta}(\theta,y')=J(\theta)- y' \mbox{ and }
\Delta'(y,y')=y-y'\]
where $S_F^\alpha(a_0) \leq y,y' \leq S_F^\alpha(b_0)$.

Now we define $V' = \phi_\theta(V)$ and
$P' = \phi_{\theta'} \circ \phi_\theta P$.
Using the conjugacy of integrals from equation (\ref{eq:conj}), equation 
(\ref{eq:difference}) becomes: 
\begin{eqnarray} \label{eq:diffinal}
V'(y,t+\tau) - V'(y,t) &=& \sum_{n=1}^
{\infty} \frac{(-1)^n}{n!}\int_{S_F^\alpha(a_0)}^{S_F^\alpha(b_0)} 
(\Delta')^n \nonumber \\
(\frac{\partial}{\partial y})^n  
\{P'(y+\Delta',t+\tau|y,t) & &  V'(y,t)\} dy'
\end{eqnarray}
Integrating over $\Delta'$, we see that $dy' = -d\Delta'$
Hence,
\begin{eqnarray}   \label{eq:Delta}
V'(y,t+\tau) - V'(y,t) &=& -\sum_{n=1}^
{\infty} \frac{(-1)^n}{n!}\int_{y-S_F^\alpha(a_0)}^{y-S_F^\alpha(b_0)} 
(\Delta')^n \nonumber \\
 (\frac{\partial}{\partial y})^n 
\{P'(y+\Delta',t+\tau|y,t) & & V'(y,t)\} d\Delta'
\end{eqnarray}
The transitional moments are given by

\begin{eqnarray} \label{eq:mn2}
\z{M}_n(y,t,\tau)=\phi_{\theta'} M_n(\theta',t,\tau) \\
=\int_{y-S_F^\alpha(b_0)}^{y-S_F^\alpha(a_0)} & & (\Delta')^n 
P'(y+\Delta',t+\tau|y,t)d\Delta'
\end{eqnarray}
Hence
\be \label{eq:7a}
M_n(\theta',t,\tau)= \int_{C(a_0,b_0)} (J(\theta)-J
(\theta'))^n P(\theta,t+\tau|\theta',t) d_F^\alpha \theta
\ee

Substituting (\ref{eq:mn2}) in equation (\ref{eq:Delta}),and applying conjugacy
 we get
\[V(\theta,t+\tau) - V(\theta,t)= \sum_{n=1}^{\infty}
 (-D_F^\alpha)^n \left\{\frac{M_n(\theta,t,\tau)}{n!}
 V(\theta,t)\right\}
 \] 
Now we assume that the moments $\z{M_n}$ and $M_n$ can be expanded in a 
Taylor series.
\be \label{eq:momentaylor1}
\frac{\z{M_n}(y,t,\tau)}{n!} = (\z{A})^{(n)}(y,t) \tau
+O(\tau^2)
\ee
and
\be     \label{eq:momentaylor}
\frac{M_n(\theta,t,\tau)}{n!} = A^{(n)}(\theta,t) \tau
+O(\tau^2)
\ee
The term of order $\tau^0$ vanishes because for $\tau=0$ the
transition probability is
$ P'(y,t|y',t) = \delta(y,y')$
which leads to vanishing moments. By taking into account only the linear terms 
in $\tau$ we have the Kramers-Moyal expansion
\be \label{eq:moyal1}
\frac{\partial}{\partial t} V'(y,t) = \sum_{n=1}^{\infty}
(-\frac{\partial}{\partial y})^n \{(\z{A})^{(n)}(y,t) V'(y,t)\}
\ee
i.e.
\be \label{eq:moyal}
\frac{\partial}{\partial t} V(\theta,t) = \sum_{n=1}^{\infty}
(-D_F^\alpha)^n \{A^{(n)}(\theta,t) V(\theta,t)\}
\ee
From the above we obtain the Fokker -Planck equation if 
the expansion in equation (\ref{eq:moyal}) stops after second term,

Thus
\begin{eqnarray} \label{eq:Fokker}
\frac{\partial V(\theta,t)}{\partial t} & = &-D_F^\alpha \{A^{(1)}(\theta
,t)V(\theta,t)\} + \\
(-D_F^\alpha)^2 & & \{A^{(2)}(\theta,t)  V(\theta,t) \}
\end{eqnarray}
where $A^{(1)}$ is the (fractal) drift coefficient and $A^{(2)}$ is the
 (fractal) diffusion coefficient.

Now we consider the special case with Gaussian transition probability 
\[P(\theta,t+\tau|\theta',t) = \frac{1}{\sqrt{\pi \tau}} \exp\{-\frac{
(J(\theta)-J(\theta'))^2}{\tau}\}\]
Then from equation (\ref{eq:7a})
\begin{eqnarray*}
M_n(\theta',t,\tau)& =& \frac{1}{\sqrt{\pi \tau}}\int_{C(a_0,b_0)}
 (J(\theta)-J(\theta'))^n \\
\exp\{-\frac{(J(\theta)-J(\theta'))^2)}{\tau} \} & &  d_F^\alpha \theta 
\end{eqnarray*}
The conjugate equation for moments gives
\[\z{M_n}(y',t,\tau) =
\frac{1}{\sqrt{\pi \tau}}
\int_{S_F^\alpha(a_0)}^{S_F^\alpha(b_0)} (y-y')^n 
\exp\{-\frac{(y-y')^2}{\tau}\} dy \]
for $S_F^\alpha(a_0) << y'$ and $ S_F^\alpha(b_0) >> y'$
we may replace the limits of the above integral by $-\infty$ and $+\infty$
respectively to get
$\z{M_1}= 0 \mbox{ and } \z{M_2} = \tau/2 $
For $n=1$ and $n=2$ using equation (\ref{eq:momentaylor1}) we see that
${\z{A}}^{(1)}=0 , {\z{A}}^{(2)} = \frac{1}{4}$ 
hence the first term on the RHS of equation (\ref{eq:Fokker}) vanishes and
$A^{(2)}$ is a constant.  We can then write equation (\ref{eq:Fokker})
as
\be \label{eq:diffusion}
\frac{\partial }{\partial t}V(\theta,t) = A (D_F^\alpha)^2 V(\theta,t)
\ee
This is a new diffusion equation on the fractal
curve $F$ with ``fractal'' diffusion coefficient $A$.

The (\ref{eq:diffusion}) is conjugate to
\[\frac{\partial V'(y,t)}{\partial t} =  A \frac{\partial^2}{\partial y^2} V'
(y,t) \mbox{ where } V'= \phi[V] \]

\begin{figure}
\resizebox{\columnwidth}{!}{\includegraphics{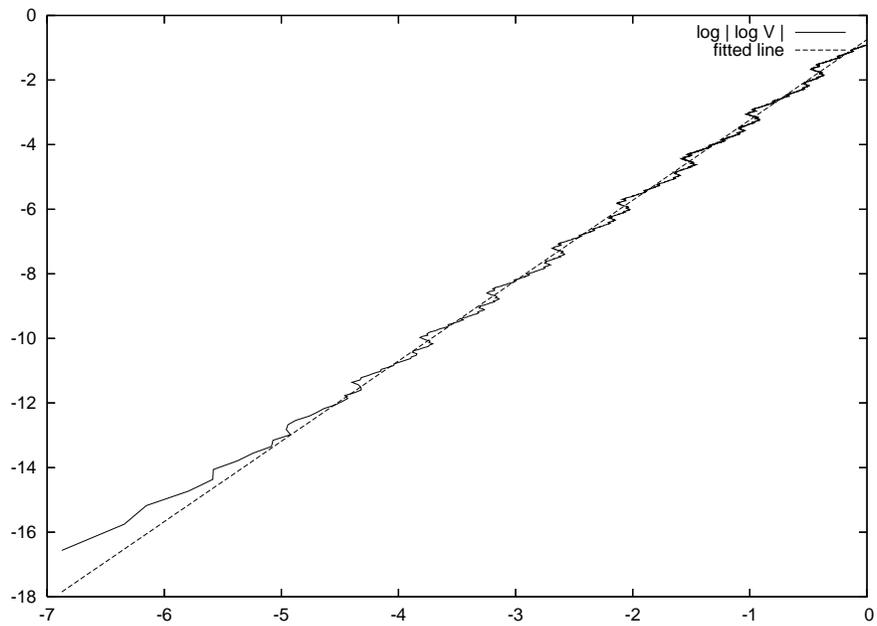}}
\caption{Plot of ($\log | \log V(\theta, t) |)$ against $(\log |\theta|)$,
for a fixed~$t$, and a straight line fit for it. The slope of the line is
2.4885, which is reasonably close to $2\alpha$, where
$\alpha = \log(4)/\log(3)=1.26$ is the dimension of the von~Koch curve.}
\label{fig:log}
\end{figure}

Given the initial condition $V'(y,0) = \delta(y)$, the solution of the above 
equation is :
\[V'(y,t) = \frac{1}{\sqrt{2\pi A t}} \exp(-\frac{y^2}{2 A t})\]
applying ${\phi}^{-1}$
\[V(\theta=\mathbf{w}(u),t) = \frac{1}{\sqrt{2\pi A t}} \exp(-\frac{(J(\theta)=
S_F^\alpha(u))^2}
{2 A t})\]
which gives the probability density $V$ at a location
$\theta = \mathbf{w}(u)$ at time $t$.
This is an exact solution of the diffusion equation~(\ref{eq:diffusion}) on a fractal curve~$F$.
The corresponding plots for the above distribution when the curve is a von-koch curve in $\mathbf{R}^2$ are shown in the
figure~\ref{fig:log}, from which the implicit
subdiffusive behaviour is clear.

\begin{figure}
\resizebox{\columnwidth}{!}{\includegraphics{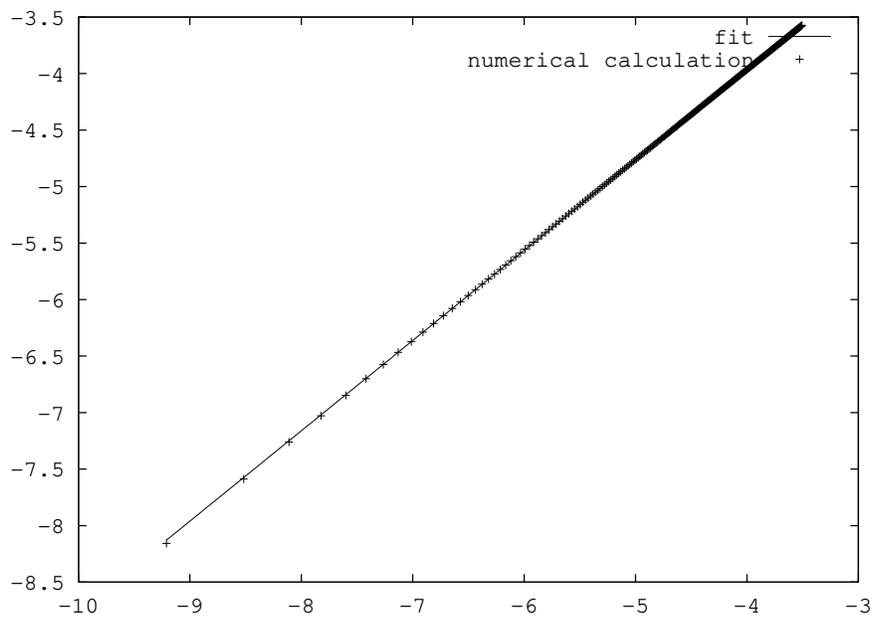}}
\caption{Plot of $log(t)$ vs $log(<L^2>)$ . The slope of the line is 0.799 
which is reasonably close to $1/ \alpha$, $\alpha = log(4)/log(3)= 1.26$,
 being the dimension of the von-koch curve.}
\label{fig:subdiffbehaviour}
\end{figure}

In figure \ref{fig:subdiffbehaviour} the subdiffusive behaviour of motion on
a fractal curve $F$ is shown. $F$ is the von-koch curve with dimension $log(4)/
log(3) = 1.26$. The relation between Euclidean distance $L(\theta)$ and time 
't' is given by 
\[\int_{C(a,b)} L(\theta)^2 P(\theta,t) d_F^\alpha \theta \sim t^{\mu}\]
where the exponent $\mu$ decides the nature of diffusion.
We find that in the above calculation, $\mu \sim 1/\alpha$, where $\alpha = 
1.26$, 
more appropriately
\be \langle L^2 \rangle \sim t^{0.802} \sim t^{1/\alpha}
\ee
and hence $\mu <1$ indicates the subdiffusive behaviour as a result of 
underlying fractal nature of space on which the particle moves.

We conclude that the underlying fractal nature of space gives rise to 
subdiffusive behaviour of the diffusing entity. It is rather a deviation
from gaussian distribution, which would have been exactly Gaussian, had the 
underlying space been ordinary and not fractal in nature. Also we see that the 
dimension of the fractal curve can be estimated from the plots of the 
distribution function.

Summarizing: We have proposed a Fokker-Planck equation on fractal curves. An
exact solution of the diffusion equation on such curves is seen to have a
subdiffusive character. The dimension of the curve can be estimated from
 the diffusion function.

\vspace{2ex}
\noindent\textbf{Acknowledgements.}
Seema Satin  is thankful to Council for Scientific and Industrial Research 
(CSIR) India, for financial assistance.

\end{document}